\documentstyle[aps,prl,epsf]{revtex}
\begin{document}
\twocolumn
\title{ Universality of Charge and Spin Response
in Doped Antiferromagnets}
\author{ P. Prelov\v sek  and J. Jakli\v c}
\address{ J. Stefan Institute, University of Ljubljana, 
61111 Ljubljana, Slovenia }
\maketitle
\begin{abstract}
Recent results for the finite-temperature static and dynamical properties
of the planar $t-J$ model, obtained by a novel numerical method for
small correlated systems, are reviewed. Of particular interest are
$T>0$ charge and spin response functions: optical conductivity
$\sigma(\omega)$ and spin susceptibility $\chi(\vec q,\omega)$, which
show very universal features at intermediate doping. In spin dynamics
the universality is best seen in the local spin correlation function,
which appears to be $T$-independent and only weakly doping
dependent. Results apply directly to the inelastic neutron scattering
and to the NMR relaxation experiments in cuprates.  Analysing
$\sigma(\omega)$ we find that the current correlation function
$C(\omega)$ is nearly $T$- and moreover $\omega$-independent in the
regime $T <J$, leading to the $\sigma(\omega)$ behavior, very close to
the marginal Fermi liquid concept and experiments.  These properties
are interpreted in connection with the large entropy, related to the large
degeneracy of the low-lying states and quantum frustration in the doped
antiferromagnets.  On the other hand, results for the uniform
susceptibility $\chi_0(T)$, as well as for the $\sigma(\omega)$, reveal
qualitative changes on entering the underdoped regime.

\end{abstract}
\section{Introduction}

Strongly correlated electrons have been in recent years intensively 
investigated theoretically, mainly in connection with the open question of 
the mechanism of superconductivity as well as the unusual metal 
properties in cuprates.  Here we will discuss only the anomalous normal 
state properties  of cuprates at $T>T_c$. These are generally attributed 
to correlation effects, nevertheless a unifying theory still remains 
a challenge.  

In this article we will address several questions related to the theoretical
understanding of the normal-state properties of cuprates:

\noindent a) Can strong correlations alone account for the unusual
normal-state properties, as manifested by the charge response: optical
conductivity $\sigma(\omega)$ and d.c. resistivity $\rho_0$, and the
spin response: dynamical spin susceptibility $\chi(\vec q,\omega)$,
related to the NMR relaxation $T_1$, etc.?

\noindent b) What is the origin of the unconventional universality within
the `normal' metallic state of cuprates, in particular in the 
`optimum-doping' regime, obtained by doping the reference antiferromagnetic
(AFM) insulator with the mobile hole-type charge carriers?

\noindent c) How many relevant energy scales exist in the doped AFM? 

\noindent d) Are the simplest prototype models, such as the $t-J$ model (and
analogous Hubbard model), enough to reproduce qualitatively and
quantitatively experimental results in cuprates?

\noindent f) Can we learn something by studying small correlated
systems at $T>0$?

From the recent experiments testing the normal-state properties
such as the d.c. resistivity \cite{batl}, optical conductivity
\cite{rome,azra}, neutron scattering \cite{shir,ster}, and the NMR and NQR
relaxation \cite{imai}, a unifying picture seems to emerge
\cite{batl}. These properties mainly depend on the extent of doping, and the
regimes have been classified into the underdoped, the `optimal'
(intermediate-doping), and the overdoped, respectively.  Whereas the
`optimal' doping refers in a narrow sense usually to samples with
the highest $T_c$, the same broader region of doping, with effective hole
concentration approx. $c_h = 0.15 - 0.25$ \cite{batl}, also exhibits
most universal properties. Down to the lowest $\tilde T\alt T_c$
resistivity follows the remarkable $\rho_0 \propto T$ law, the uniform
susceptibility $\chi_0$ is varying monotonously with $T$, etc. On the
other hand, the underdoped materials (effective $c_h \alt 0.15$) are
characterized by the onset of the `pseudogap' at $T<\tilde T\gg T_c$, which
shows up in the kinks in $\rho(T)$, $\chi_0(T)$ etc.

We investigate in this review mainly the intermediate-doping (optimal)
regime.  We discuss finite-$T$ (normal-state) properties of the doped AFM
within the $t-J$ model \cite{rice}
\begin{equation}
H=-t\sum_{\langle ij\rangle  s}(c^\dagger_{js}c_{is}+ H.c.)
+J\sum_{\langle ij\rangle} (\vec S_i\cdot \vec S_j - {1\over 4} n_i n_j ),
\label{eq1}
\end{equation}
where $c^\dagger_{is}(c_{is})$ are projected fermionic operators,
prohibiting double occupancy of sites. The ground state of the $t-J$
model and related properties have been intensively studied both by
analytical \cite{rice} and numerical methods \cite{dago}, nevertheless
some basic questions, e.g. concerning possible pairing in the ground
state and the origin of  anomalous metallic properties at finite $T>0$, 
remain unsolved.

\section{Fermi liquids}

Normal metals are expected to follow the concept of the {\bf Landau Fermi
liquid}.  The excitations in this case should be quasiparticles with
well defined $\vec q$ and energy $\epsilon_{\vec q}$ near the Fermi
energy, with the damping $\tau(\vec q)^{-1} \propto (\epsilon_{\vec q}
- E_F)^2$. Consequences are well known, e.g. for the dynamical
susceptibility
\begin{equation}
\omega \rightarrow 0:~~\chi''(\vec q,\omega)/\omega \ne f(T),
\label{eq31}
\end{equation}
leading to the Korringa law ${\mathrm{NMR}}~~T_1^{-1} \propto T $.
Also, the Drude form is expected for $\sigma(\omega)$ with 
$\tau^{-1} \propto T^2$ etc. 

The normal-state properties of cuprates, however, do not follow the Landau
scenario. Experiments have been mainly described within two alternative
concepts. The {\bf nearly antiferromagnetic Fermi liquid} has been introduced
by Moriya and Millis et al. \cite{mill} to describe the spin
fluctuations in the doped AFM, mainly in connection with the anomalous NMR
relaxation phenomena in cuprates. The low-$\omega$ spin response is
modeled with the overdamped spin fluctuations peaked at the AFM
wavevector $\vec Q =(\pi,\pi)$,
\begin{equation}
\chi''(\vec q, \omega) = {\omega \chi_0 \Gamma \xi^2 \over
\omega^2 + \Gamma ^2 [1+|\vec q-\vec Q|^2 \xi^2 ]^2} .\label{eq32}
\end{equation}
AFM correlation length $\xi(T)$ is assumed to be critical in the doped
regime, in particular with $\xi^2(T) \propto 1/T$ in the $z=2$
dynamical RPA-type theory \cite{mill}, while the postulated mapping to
the quantum critical regime of the nonlinear sigma model would imply
$\xi(T) \propto 1/T$ \cite{soko}.

Alternative scenario has been summarized within the {\bf marginal
Fermi liquid} (MFL) hypothesis \cite{varm}. Here, the spin and charge
susceptibility have been postulated in the anomalous form
\begin{eqnarray}
\chi''(\vec q, \omega) &\sim& C ~{\omega\over T},~~~|\omega|<T, 
\nonumber \\
\chi''(\vec q, \omega) &\sim& C~ {\mathrm {sgn}}  \omega, ~~~|\omega| >T, 
\label{eq33}
\end{eqnarray}
leading to the marginality of the quasiparticle self energy
\begin{equation}
\Sigma(\omega) \sim \lambda \lbrack 2 \omega ~{\mathrm{ln}} 
{\pi T - i \omega \over \omega_c} - i \pi ^2 T \rbrack. \label{eq34}
\end{equation}
Consistent with Eq.(\ref{eq34}), the low-frequency $\sigma(\omega)$
has been written in the generalized Drude form introducing an
anomalous effective relaxation rate $\tau^{-1} \sim 2\pi \lambda
(\omega + \pi T)$.

\section{Finite temperature method}

$T>0$ properties of the $t-J$ model have been so far calculated only
by the high-temperature series expansion \cite{sing} and via full or
partial diagonalization of quite small systems. Recently the present
authors introduced a novel numerical method, based on the Lanczos
diagonalization method combined with random sampling
\cite{jakl}. The latter method has the advantage that it allows the
study of dynamical and static response functions within the most
challenging regime $T,\omega < J$, where one could hope for a
universal behavior, as observed in the experiments on cuprates. It
should be noted that analogous Hubbard model seems more feasible for
$T>0$ studies, due to the possible application of the Quantum Monte Carlo
methods \cite{dago}. Nevertheless, in the doped case the well known minus
sign problem so far mainly prevented  calculations within the
regime $T,\omega <J$ \cite{preu}.

Let us present briefly our method for $T>0$ calculations, showing in
more detail the evaluation of the statical expectation value of an
operator $A$,
\begin{equation}
\langle A \rangle= Z^{-1}\sum_n^N \langle n|{\mathrm e}^{-\beta H}
A|n\rangle, ~~~
 Z= \sum_n^N \langle n|{\mathrm e}^{-\beta H}|n\rangle, \label{eq51}   
\end{equation}
where $\beta=1/T$ (we use $k_B=\hbar=1$ furtheron) and the sum runs
over the chosen complete basis set of orthonormal wavefunctions
$|n\rangle, n=1,N$, spanning the Hamiltonian $H$. If we could perform
the full diagonalization of the problem and find all eigenstates
$|\Psi_l\rangle$ and corresponding energies $E_l$, we would express
the result in a usual way,
\begin{equation}
\langle A \rangle= \sum_l^N {\mathrm e}^{-\beta E_l} 
\langle l|A|l\rangle / \sum_l^N {\mathrm e}^{-\beta E_l}. \label{eq52}
\end{equation}

We choose an alternative approach. With each basis function
$|n\rangle$ we start a Lanczos procedure  $|\phi_0^n\rangle
=|n\rangle$, generating an orthonormal set of functions
$|\phi_m^n\rangle, m=0, M$,
\begin{eqnarray}  
H|\phi^n_0\rangle &=& a_{n0} |\phi^n_0\rangle +b_{n1}|\phi^n_1\rangle,
\nonumber \\
H|\phi^n_m\rangle &=&b_{n m}|\phi^n_{m-1}\rangle + a_{nm} 
|\phi^n_m\rangle 
+b_{n m+1}|\phi^n_{m+1}\rangle, \\ \label{eq53}
H|\phi^n_M\rangle &=& b_{n M} |\phi^n_{M-1}\rangle +a_{n M}|\phi^n_M
\rangle. \nonumber 
\end{eqnarray}
For the chosen number of Lanczos steps $M \ll N$ we then diagonalize 
the tridiagonal matrix of coefficients $a_{nm}, b_{nm}$ to find the 
energies $\epsilon_{nm}$ and the corresponding eigenfunctions 
$|\psi^n_m\rangle$. Within the restricted basis of these functions one
can write the approximation for Eq.(\ref{eq52}) as 
\begin{eqnarray}
\langle A \rangle&=&Z^{-1}\sum_n^{N_0} \sum_m^M\langle 
n|\psi_m^n\rangle {\mathrm e}^{-\beta \epsilon_{nm}}
\langle\psi_m^n|A|n\rangle, \nonumber \\
Z&=& \sum_n^{N_0} \sum_m^M|\langle
n|\psi_m^n\rangle|^2{\mathrm e}^{-\beta \epsilon_{nm}}. \label{eq54} 
\end{eqnarray}
It is evident that the average evaluated via Eq.(\ref{eq54}) is equivalent
to Eq.(\ref{eq52}) for the sampling over the full basis set $N_0=N$ and 
$M=N-1$. 

Our claim is that very accurate results can be obtained via
Eq.(\ref{eq54}) even for a severly reduced number of Lanczos steps
$M\ll N$ and for a partial random sampling of the basis states $N_0\ll N$,
instead of the full sampling. Let us first investigate the method for
the full sampling $N_0=N$. We note that the Lanczos
procedure Eq.(\ref{eq53}) represents an iterative action of the
operator $H$ on the initial function $|n\rangle$. Performing the (high
temperature) expansion in $\beta$ of the numerator and of the
denominator of Eq.(\ref{eq54}), it is easy to prove that both power
series are correct up to the order $M$. On the other hand, 
$\langle A\rangle$ evaluated via Eq.(\ref{eq54}) is
accurate also for $T=0$ (provided $M>M_0$), since it is a standard
experience that the Lanczos procedure converges quite rapidly,
typically in $M=M_0\sim 50 \ll N$ steps, from an arbitrary initial
function $|n\rangle$ (not orthogonal to the ground state) to the
ground state energy $E_0$ and the corresponding wavefunction
$|\Psi_0\rangle$.  Hence, at full sampling our Lanczos-based method
represents an optimum interpolation scheme between the high
temperature $T \rightarrow \infty$ and the ground state $T=0$
result.

The second important ingredient is the reduction of the sampling to
the partial random one with $N_0\ll N$. This step cannot be justified
rigorously, but seems plausible in analogy with the statistical Monte
Carlo methods. The most severe test is expected to be the result at
$T=0$, which can be anyhow tested via an alternative direct evaluation
with the known $|\Psi_0\rangle$. The general experience is that good
convergence is achieved for $N_0\ll N$.

Even more appealing is the application of the method to the calculation
of dynamical quantities. We choose as an example the optical conductivity,
expressed within the linear response theory as 
\begin{eqnarray}
 \sigma(\omega)&=&{1-{\mathrm e}^{-\beta\omega}\over \omega} 
\mathrm{Re}\int^{\infty}_{0} dt~ {\mathrm 
e}^{i\omega t} C(t) , \nonumber \\ C(t) = \langle j(t) j\rangle &=&
Z^{-1} \sum_n\langle n|{\mathrm e}^{(-\beta +it)H} j {\mathrm
e}^{-iHt}j |n\rangle.  \label{eq56}
\end{eqnarray}
In order to get the result at arbitrary $\beta$ a double sum over all 
eigenstates $|\Psi_l\rangle, |\Psi_{l'}\rangle$ is required.
Instead, we use the approach described above for $\langle A\rangle$ 
generalized to dynamical $C(t)$,
\begin{eqnarray}
C(t)= Z^{-1} \sum_n^{N_0}  \sum_{m,k}^M && \langle n|\psi_m^n\rangle
{\mathrm e}^{-\beta \epsilon_{nm}} {\mathrm e}^{i (\epsilon_{nm} -\tilde
 \epsilon_{n k})t} \times \nonumber \\
&&\langle\psi_m^n|j|\tilde \psi_k^n \rangle \langle
\tilde \psi_k^ n |j| n\rangle. \label{eq58} 
\end{eqnarray}
Here,  $|\tilde \psi_k^n \rangle$ and $\tilde \epsilon_{n k}$ are
generated by the Lanczos procedure analogous to Eq.(\ref{eq53}) via the 
orthonormal basis set $|\tilde \phi_k^n\rangle $, but with the 
initial condition
\begin{equation}
|\tilde \phi_0^n\rangle= j|n\rangle / \sqrt{\langle n|j^2|n \rangle}. 
\label{eq59}
\end{equation}
While for $N_0=M=N$ the expression (\ref{eq58}) is equivalent to the
exact one, the arguments for using $N_0\ll N$ and $M\ll N$ are analogous
to those described above for the static case. E.g., for the full
sampling $N_0=N$ Eq.(\ref{eq58}) yields correct series in $-\beta+it$
and $it$, respectively, up to the $M^{\mathrm th}$ order. Accordingly,
$\sigma(\omega)$ has correct frequency moments $\langle \omega^p
\rangle,~~p=0,M$, for $\beta\rightarrow 0 $ etc.
Again for $M>M_0$ the method yields meaningful result also for $T=0$,
which can be tested via the usual Lanczos method for the $T=0$ dynamical
quantities \cite{dago}.

It is important to stress that computational requirements of the
method for a system of the given size are comparable to those for the ground
state evalutions via the Lanczos method, whereby the CPU time clearly
increases with the number of random samples $N_0$. The CPU time and
memory requirements are also enhanced due to the necessary
reorthogonalization of Lanczos functions and due to the evaluation of
matrix elements in Eq.(\ref{eq58}).

Let us finally comment on the finite size effects, which due to the
restricted system sizes still remain the main obstacle to achieve
reliable results for low $T$. It is a plausible observation that the
method yields macroscopic-like results for $T>T^*$, where $T^*$ is
roughly determined by the average level separation in the low-energy
many-body spectrum, being dependent on the system size.  For $T<T^*$
the finite-size effects become pronounced, e.g. dynamical quantities 
in general reveal (spurious) peak structures etc. It is however very
important to note that within the $t-J$ model, the $T^*$ is smallest within
the intermediate (optimum) doping regime $2/16 \le c_h \le 4/16$,
where we typically reach $T^* \sim 0.1~t$ for systems $N=16,18$. 
The $T^*$ is larger both within the underdoped and the overdoped
region (for the fixed system size).
 
\section{Spin Susceptibility}

Let us consider the dynamical spin response, as given by the
susceptibility $\chi(\vec q, \omega)$ and the corresponding
dynamical spin correlation function $S(\vec q,
\omega)$
\begin{eqnarray}
\chi''(\vec q, \omega) &=&  (1 - \mathrm e^{-\beta \omega})
S(\vec q, \omega), \nonumber \\
 S(\vec q, \omega) &=& \mathrm {Re}\int_0^{\infty} dt~ \mathrm e^{i \omega t }
\langle S^z_{\vec q}(t) S^z_{- \vec q} \rangle . \label{eq2}
\end{eqnarray}
It has been shown by the present authors \cite{jakl2} that 
$\chi''(\vec q,\omega)$ shows the coexistence of the high-frequency 
($\omega \propto t$) free-fermion-like
contribution and the low-$\omega$ spin-fluctuation contribution. At
the same time, $\chi''(\vec q, \omega<T)$ reveals a pronounced
$T$ dependence, consistent with the MFL form, and (even
quantitatively) with the NMR relaxation in cuprates.

It appears quite helpful to concentrate on the local spin correlation
function $S_L(\omega)$ \cite{jakl3} and its symmetric part $\bar
S(\omega)$
\begin{eqnarray}
S_L(\omega)&=&{1\over N} \sum_{\vec q} S(\vec q, \omega),\nonumber\\
\bar S(\omega)=S_L(\omega)&+&S_L(-\omega)= (1+{\mathrm 
e}^{-\beta\omega})  S_L(\omega). \label{eq3}
\end{eqnarray}
It should be pointed out that $S_L(\omega)$ and the related
susceptibility $\chi_L(\omega)$ are directly measured (in cuprates) by
neutron scattering
\cite{shir,ster}. The NMR relaxation as well yields the information on
$S_L(\omega \rightarrow 0)$ \cite{imai}, provided that the AFM spin
fluctuations $\vec q \sim \vec Q=(\pi,\pi)$ are dominant. 

An important restriction for $\bar S(\omega)$ is the sum rule
\begin{equation}
\int_0^{\infty}  \bar S(\omega) d\omega = \pi \langle (S_i^z)^2\rangle = 
{\pi \over 4} (1-c_h), \label{eq4}
\end{equation}
where $c_h=N_h/N$ is the hole concentration. 

We perform the evaluation of $\bar S(\omega)$ via
Eq.(\ref{eq3}) by calculating $S(\vec q,\omega)$, using the finite-$T$
diagonalization method for small systems, in this case for the $t-J$
model on the square lattice with $N=16, 20$ sites. We fix $J/t =
0.3$ to remain in the regime of cuprates \cite{rice}. 
We stress again that the method yields macroscopic-like results for
$T>T^*\agt 0.1~t$.

In Fig.~1 we display the $\bar S(\omega)$ for $c_h=1/20, 3/16$ and
several $T$ in the range $0.1 < T/t <0.7 $ \cite{jakl3}. It is
immediately evident that $\bar S(\omega)$ at `optimal' doping
$c_h=3/16$ is essentially $T$-independent in a wide $T$-range,
although one crosses the exchange-energy scale $T \sim J$.  For the
underdoped case $c_h=1/20$ the behavior is analogous for higher $T
>T_0 \sim 0.7~J$, consistent with the quantum critical regime
within the AFM
\cite{chac}. Deviations at lower $T< T_0$ (where the renormalized
classical regime \cite{chac} is expected in the AFM) could be an
indication for the onset of a `pseudogap'.

\begin{figure}
\epsfxsize=8cm
\epsffile[70 -600 670 -50]{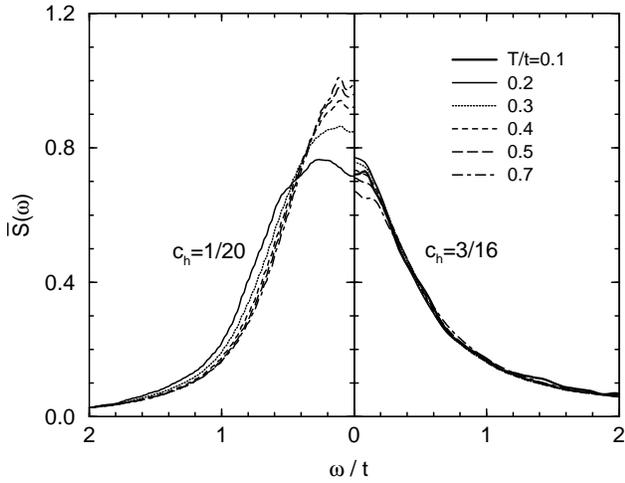}
\caption{Local spin correlation function $\bar S(\omega)$ for
 $c_h$=1/20, 3/16 and various $T$.}
\end{figure}

To follow the doping
dependence we present in Fig.~2 the variation with $c_h$ at fixed
$T=0.2~t<J$, plotting the integrated intensity
\begin{equation}
I_S(\omega)= \int_0^{\omega}  \bar S(\omega') d\omega'. \label{eq5}
\end{equation}
Again, we notice that for the chosen $T$ the results are most 
reliable at the intermediate doping.
The most striking message is that the initial slope
of $I_S(\omega)$ and consequently $S_L(\omega \rightarrow 0)$ are
nearly doping independent for $0\le c_h \le 0.25$.
This is well consistent with the NMR
(NQR) relaxation rates $T_1^{-1}$ as measured in $\mathrm{La_{2-x}
Sr_x Cu O_4}$ in the range $x=0 - 0.15$
\cite{imai,jakl2}.

Only for the overdoped systems with $c_h>0.25$ the low-frequency behavior
changes qualitatively, where the latter part is strongly suppressed as
expected in (more) normal Fermi liquids. $I_S(\omega>J)$ is doping
dependent even for $c_h<0.25$, consistent with the $c_h$-dependence of
the sum rule, Eq.(\ref{eq4}). In addition, at the intermediate doping
$\bar S(\omega)$ decreases smoothly (see Fig.~1) up to $\omega
\sim 4t$, this being the consequence of a free-fermion-like component
\cite{jakl2}. On the other hand, in the underdoped regime the dynamics 
is restricted to $\omega <3J <t$. 

\begin{figure}
\epsfxsize=8cm
\epsffile[30 -600 645 -50]{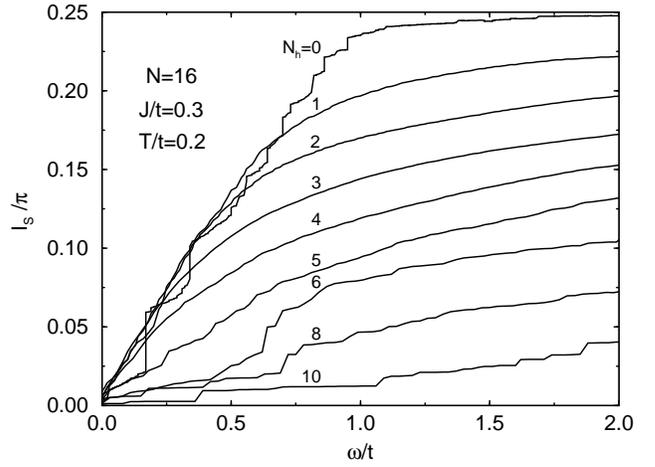}
\caption{Integrated spectra $I_S(\omega)$ at fixed $T=0.2~t$ and 
various $c_h$.}
\end{figure}

How can one explain the universality of $\bar S(\omega)$ at
intermediate $c_h$? First, we note that up to $c_h \sim 0.3$ the
dominant scale of spin fluctuations remains related to $J$. From the
explicit expression in terms of the eigenstates of the system
\begin{eqnarray}
\bar S(\omega) = (1+\mathrm e^{-\beta\omega}) {\pi \over Z}
\sum_{n,m}  \mathrm e^{-\beta E_n} |\langle n| S_i^z|m \rangle|^2  
\times \nonumber\\
\delta (\omega - E_m+E_n), \label{eq6}
\end{eqnarray}
one would conjecture (quite generally) a $T$-independence of response
for $\omega \gg T$. Although such plausibility arguments have been
used previously, their validity clearly depends on the character and
on the density of low-lying many-body states. To explain also the 
$T$-independence of $\bar S(\omega<T)$, we only need to recall the 
sum rule Eq.(\ref{eq4}) and to assume that there is no characteristic scale 
$\omega_c< T$,
which could introduce an additional low-$\omega$ structure in $\bar
S(\omega)$.

A natural scale for the AFM is the (gap) frequency $\omega_c \sim c /\xi$,
where $c$ is the spin wave velocity and $\xi$ the AFM correlation length. 
Here originates the essential difference between the undoped and the
`optimally' doped AFM. While for an AFM in the renormalized classical
regime $\xi$ is exponentially large for $T\ll J$, and consequently
$\omega_c < T$ \cite{chac}, in the doped case $\xi < 1/\sqrt{c_h}$ is
determined predominantly by $c_h$, so $\xi$ is rather $T$-independent 
for $T<J$, excluding $\omega_c<T<J$. In other words, the observed
anomalous low-$\omega$ behavior within the $t-J$ model does not
originate in a strong $T$-dependence of $\xi$, as conjectured by the
nearly AFM Fermi liquid, Eq.(\ref{eq32}).

We conclude the discussion of spin dynamics by the consequences of the 
universality of $\bar S(\omega)$.  The  local susceptibility is given by
\begin{equation}
\chi''_L(\omega) = \mathrm {tanh}\left( {\omega \over 2T} 
\right) \bar S(\omega). \label{eq7}
\end{equation}
Since neutron scattering probes only $\omega<J$, one can simplify
Eq.(\ref{eq7}) further by $\bar S(\omega) \sim \bar S_0$.  Such form
has been recently used to describe experiments \cite{ster}. It is also
qualitatively consistent (but as well more restricted in the form) with
the MFL Ansatz, Eq.(\ref{eq33}) \cite{varm}.

\section{Optical conductivity}

Let us investigate in an analogous way the dynamical conductivity
$\sigma(\omega)$, Eq.(\ref{eq56}).  Results for this quantity by the
present authors \cite{jakl1}, obtained by the finite-$T$ method, can
be summarized as follows: a) in the intermediate-doping regime
$\sigma(\omega)$ shows a non-Drude fall-off, consistent with the
$\omega$-dependent relaxation rate within the MFL concept
Eq.(\ref{eq34}) \cite{varm}, deduced from experiments in cuprates
\cite{rome,azra}, and b) qualitative as well as quantitative results 
for $\sigma(\omega)$ and $\rho(T)$ agree well with the
experimental ones.

Unlike $\bar S(\omega)$, $C(\omega)$ does not obey a $T$-independent
sum rule.  Nevertheless, motivated by the universal spin dynamics we
reexamine in an analogous manner our results on $\sigma(\omega)$
\cite{jakl1}.  In Fig.~3 we present the corresponding integrated
spectra $I_C(\omega)=\int_0^{\omega} C(\omega') d\omega'$ at fixed
doping in the `optimal' regime for various $T\le t$ \cite{jakl3}. We
establish that for $T \le J$ spectra $I_C(\omega)$ are essentially
independent of $T$, at least for the available $T>T^*$.  At the same time
the slope of $I_C(\omega < 2~t)$ is nearly constant, i.e.  $C(\omega)
\sim C_0$ in a wide $\omega$-range, $C_0$ being weakly $J$-dependent
(tested for $J/t=0.2, 0.6$).

We note that such $C(\omega) \sim C_0$ implies a nonanalytic behavior
of $\sigma(\omega \rightarrow 0)$, starting with a finite slope at
$\omega=0$.  Moreover, with $C(\omega)=C_0$ we can claim a simple
universal form for $\omega<2~t$
\begin{equation}
\sigma(\omega)=C_0 {1-\mathrm e^{-\beta\omega} \over \omega}.
\label{eq10}
\end{equation}
At first sight, it appears rather surprising that this form can be
well fitted (for $\omega, T\ll t$) with a Drude-type form with an MFL-like
effective relaxation rate $\tau^{-1}=2\pi \lambda (\omega + \eta T)$
with specific $\lambda \sim 0.09$ and $\eta
\sim 2.7$ \cite{jakl1}.  The form Eq.(\ref{eq10}) for $\sigma(\omega)$
trivially reproduces the remarkable linear law $\rho \propto T$ in
cuprates \cite{batl}, as well as the non-Drude falloff at $\omega >T$.

\begin{figure}
\epsfxsize=8cm
\epsffile[20 -600 650 -50]{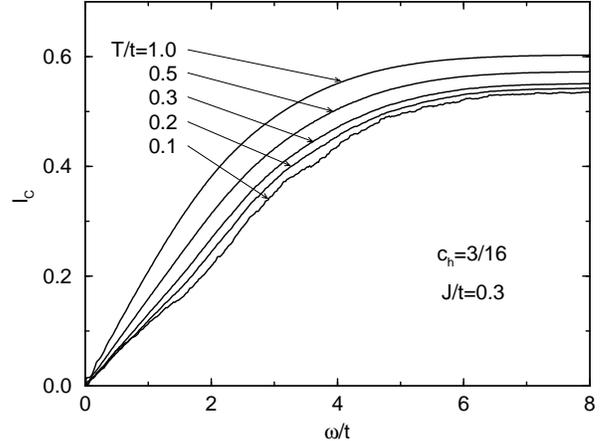}
\caption{Integrated current-correlation spectra $I_C(\omega)$ for
$c_h=3/16$ and various $T$.}
\end{figure}

A quantitative comparison (assuming $J/t=0.3$ and $t=0.4~\mathrm{eV}$) of
$\sigma(\omega)$ with the experimental results \cite{rome,azra} is quite
remarkable \cite{jakl1}, realizing the simplicity of the model.  It is
however evident that the expression (\ref{eq10}) is more restricted
than the MFL Ansatz, hence it should be retested in more detail with
experiments.

It should be mentioned that $C(\omega)\sim C_0$ has been derived for a
single hole conductivity within the retraceable path approximation
\cite{rice1}, with a restricted validity for $T>t$ (or possibly $T>J$).
We find this behavior only for the intermediate doping, hence new
arguments are needed. We can follow the analysis analogous to $\bar
S(\omega)$, Eq.(\ref{eq6}), expressing $C(\omega)$ in terms of the
eigenstates. As before, the $T$-independence of $C(\omega>T)$ seems
plausible. The fact that $C(\omega<2t)\sim C_0$ requires, however, that
the current relaxation is very fast, i.e. determined only by the
incoherent hopping and by the interhole collisions. The spin fluctuation
scale $J$ does not enter directly, i.e. even for $T\ll J$ the spin
system only serves as a random bath for the charge degrees of freedom
(holons). This conclusion remains valid as far as there is no characteristic
frequency $\omega_c<T$ (e.g. the `pseudogap') in the system.

On the other hand, our results also indicate the possible qualitative
changes when entering the underdoped regime.  E.g. the $c_h=2/18$ doping
seems to represent a crossover to the latter behavior, where in
contrast (e.g.  for $c_h=1/20$) we find a pronounced $T$- and
$\omega$-dependence of $C(\omega)$ at $T\le J$.  This qualitative
change can be again attributed to a `pseudogap' appearing at larger
$T$ in the underdoped systems \cite{batl}.

\section{Entropy}

Above arguments, both for $S_L(\omega)$ and $C(\omega)$, require a
large density (degeneracy) of the low lying many-body states, apparently
being a crucial feature for the most challenging intermediate-doping
regime. To quantify this statement we calculate the entropy density
$s=S/N$, using again the finite-$T$ diagonalization method, being less
space and time consuming for static quantities \cite{jakl}. In Fig.~4
we present the results for $s(c_h)$, obtained for $N=16$ and $N=20$ at
different $T$. The main lesson from Fig.~4 is that the `optimal' cases
with $c_h \sim 0.15-0.3$ are characterized by the largest entropy $s$
at low $T<J$, e.g.  $s > 0.2/$site for $T \sim 0.2~J$, being almost
one half of $s(T=\infty)$ for an AFM. This implies a very large
degeneracy of the low-lying states, which could be attributed to the spin
subsystem, frustrated by the hole motion. Calculated entropy, large
even at moderate $T$ and strongly dependent on doping, seems to be
(also quantitatively) consistent with the specific heat measurements on
cuprates
\cite{lora}.

\begin{figure}
\epsfxsize=8cm
\epsffile[40 -600 620 -50]{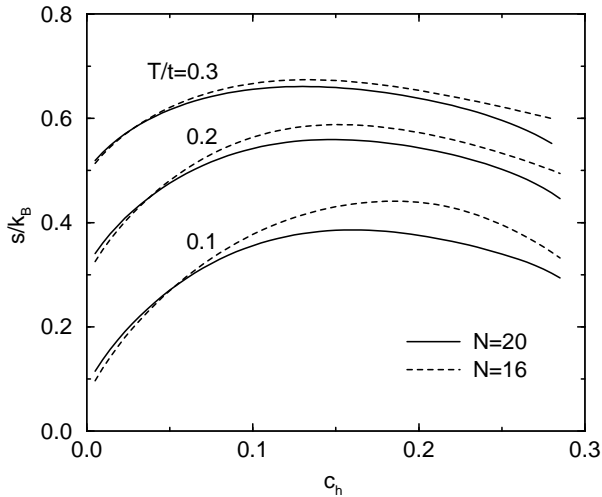}
\caption{Entropy density $s$ vs. $c_h$ for three $T$ and two system
sizes $N$.}
\end{figure}

\section{Conclusions}

Our results for the $t-J$ model seem to indicate that
the anomalous, but universal, dynamics in the intermediate-doping
regime of correlated systems is a consequence of an extreme
degeneracy and a collapse of the low-lying quantum states (introduced by
doping the magnetic insulator). This leads to a diffusive-like charge
and spin response, where $T$ represents the only relevant energy
(frequency) scale.  On the other hand, our results do not exclude the
possible onset of coherence (related to `pseudogaps' and
superconductivity) at lower $T<T^*\sim 0.1~t$.

It is evident that the `optimal' doping regime is characterized by the
competing influence of two quantum processes: AFM fluctuations and
hole hopping.  Outside this regime, either one of the two processes
should dominate. 

\begin{figure}
\epsfxsize=8cm
\epsffile[40 -600 620 -50]{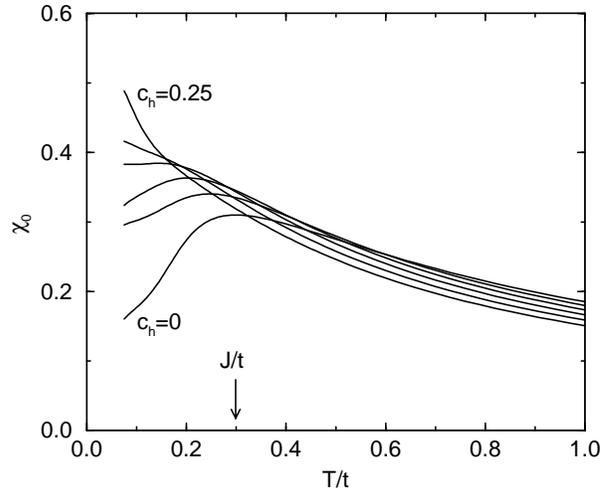}
\caption{Uniform susceptibility $\chi_0$ (in units of  $1/t$) vs. $T$ 
for various $c_h$ (increasing in steps of 0.05), 
as obtained by calculations on the system with $N=20$ sites
for $c_h>0$, and $N=26$ sites for $c_h=0$, respectively.}
\end{figure}

To show that such physics is indeed incorporated within
the $t-J$ model, we present our results for the uniform susceptibility
$\chi_0(T)$ at various hole dopings $c_h$.  From Fig.5 we establish
that $\chi_0(T)$ is most monotonous within the intermediate-doping
regime. On the other hand, for $c_h <0.15$ the susceptibility starts
to reveal a shoulder at $\tilde T$, which could be interpreted as an
increase of the pseudogap scale, but as well as a sign of the dominating
role of the AFM fluctuations.

\end{document}